\documentclass[final,5p,times,twocolumn]{elsarticle} 
\usepackage[utf8]{inputenc}

\usepackage{hyperref}

\journal{}
\usepackage{eurosym}
\usepackage{amsmath, amssymb, setspace}
\usepackage{subcaption}

\begin{document}

\begin{frontmatter}

\title{PETALO: a Time-of-Flight PET with liquid xenon}
\author{C. Romo-Luque, on behalf of the PETALO Collaboration}
\address{Instituto de Física Corpuscular (IFIC), CSIC \& Universitat de València
	
Calle Catedrático José Beltrán, 2, 46980 Paterna, Valencia, Spain}

\begin{abstract}

Liquid xenon offers several features, which make it suitable for applications in nuclear medicine, such as high scintillation yield and fast scintillation decay time. Moreover, being a continuous medium with a uniform response, liquid xenon allows one to avoid most of the geometrical distortions of conventional detectors based on scintillating crystals. In this paper, we describe how these properties have motivated the development of a novel concept for positron emission tomography scanners with Time-Of-Flight measurement, which combines a liquid xenon scintillating volume and silicon photomultipliers as sensors. A Monte Carlo investigation has pointed out that this technology would provide an excellent intrinsic time resolution, down to 70 ps. Also, the transparency of liquid xenon to UV and blue wavelengths opens the possibility of exploiting both scintillation and Cherenkov light for a high-sensitivity positron emission tomography scanner with Time-Of-Flight capabilities. Monte Carlo simulations point to a time resolution of 30-50 ps obtained using Cherenkov light. A prototype is being built to demonstrate the high resolution in energy, time and reconstruction of spatial coordinates of this concept, using a ring of 30 cm internal diameter and a depth of 3 cm instrumented with VUV-sensitive silicon photomultipliers.
\end{abstract}


\end{frontmatter}

\section{Positron Emission Tomography}

 Positron Emission Tomography (PET) is a noninvasive technique that produces 3D images of metabolic processes in the body as an aid to the diagnosis of diseases. The principle of operation is the injection of a biologically active molecule, an analogue of glucose doped with a radioactive positron emitter isotope, into the patient. The molecule is distributed inside the patient via the blood flow and accumulates in tissues and organs with high metabolism. The emitted positron annihilates with an electron from its surroundings giving rise to two 511 keV photons which can be detected in coincidence by a ring of detectors. Each coincidence detection provides a Line of Response (LOR). From the detection set of all the LORs, an image of the radiotracer concentration is reconstructed.

This technology is a medical and research tool, mainly used for diagnosis, staging and monitoring treatments of cancer as well as brain imaging and diagnosing disorders of the central nervous system. However, blood flow, heart related diseases and any metabolic process can be studied with specific radiotracers.

True events, in which the two photons are detected within a temporal window of few ns, are said to be prompt or coincidence events. Events in which at least one of the photons suffers a Compton scattering inside the patient and events where the two detected photons originate from different annihilations, produce an incorrect LOR and can be reduced by using an energy window centered on the 511 keV photopeak and improving the resolution in time.

\section{Time of Flight}
The concept of time of flight (TOF) refers to the precision in the measurement of the arrival time of the two photons. Since the emission of both photons is simultaneous, the difference in their arrival times provides the difference in the path lengths of the photons. Therefore, the location of the emission position can be inferred and is given by a gaussian distribution.

Coincidence Resolving Time (CRT) is the parameter determining noise reduction in TOF-PET and it is calculated by measuring the FWHM of the distribution of TOF measurements. As the timing resolution enhances, the signal-to-noise ratio of the image will improve, requiring fewer events to achieve the same image quality and thus shorter reconstruction time and lower dose for the patient.

\section{Liquid xenon}

The basic unit employed in PET detectors is a scintillation crystal coupled to a photodetector, either a photomultiplier tube or an avalanche photodiode. Most modern PET scanners are based on LSO or LYSO, which combine high density and high yield scintillation.

Liquid xenon (LXe) has been proposed as a detector medium for medical applications since the 1970’s. Lavoie introduced the idea of using LXe for a PET scanner in 1976 \cite{doi:10.1118/1.594289}. In the following decades, several approaches have been studied, such as measuring scintillation and ionization in LXe with Time Projection Chambers \cite{chepel} or performing calorimetry in LXe scintillators read out by photomultipliers \cite{Doke:1999ku}. 

Liquid xenon has excellent scintillation features that can improve the performance of a PET scanner using TOF. It has a large scintillation yield of 68 photons per keV and a fast scintillation of 2.2 ns for the first component, compared to LSO (LYSO) that has a scintillation yield of 32 photons per keV and a decay time of 40 ns. Since its state is liquid and its density is high, a compact homogeneous detector with uniform response that minimizes dead regions can be built. Also, a detector made of a continuous liquid target is easier to scale than a detector made of many discreet crystals. At atmospheric pressure xenon liquifies at 161.35 K and thus the required cryogenics are relatively simple. Furthermore, operation at this temperature reduces the dark noise of sensors such as SiPMs to negligible levels. 

In addition, LXe emits Cherenkov light, whose promptness is a few picoseconds to be compared with hundreds of picoseconds for scintillation light and can be exploited for TOF measurements. On the other hand, LXe is transparent to UV and blue light and Cherenkov and scintillation light can be distinguished due to the difference in wavelength:  178 nm for scintillation in comparison with around 300 nm for the fastest Cherenkov photons. Also the low cost of this material is remarkable compared to crystals. 

\section{The PETALO concept}

PETALO (Positron Emission TOF Apparatus based on Liquid xenOn) is a new concept that seeks to demonstrate that liquid xenon together with a SiPM-based readout and fast electronics, provide a significant improvement in the field of medical imaging with PET-TOF \cite{Ferrario:2017sgq}.

PETALO uses exclusively LXe scintillation light, which provides enough information to achieve good energy and spatial resolution (of the same order than that achieved by conventional LYSO scanners), plus a high resolution measurement of CRT. While the possibility of reading also the ionization charge has been explored by several groups \cite{gallegomanzano:in2p3-01166249} \cite{articleChepel} and it has been shown that the combination of ionization charge and scintillation light results in better energy and spatial resolution, such detectors do not have good TOF resolution and are very slow due to the long times needed to drift the ionization charge. 

The detector consists of a continuous ring surrounded by a dense array of VUV SiPMs and can be optimised depending on the intended application. SiPMs enable fine granularity for high spatial resolution in the reconstruction of the gamma interaction, and they have good photodetection efficiency (PDE) and low dark count rate at cryogenic temperatures. Either the entry or the exit face, or both, may be instrumented with SiPMs, while the uninstrumented faces can be covered with an absorber material in order to avoid reflections and spatial distortions.
The ring is located inside a cylindrical vacuum chamber with room temperature bore, leaving a cavity where the patient will be placed.

Monte Carlo (MC) studies were performed using Geant4 and a simple geometry: two 5 $\times$ 5 $\times$ 5 cm$^{3}$ cells with the entry and the exit faces instrumented. Between them, in the centre of the geometry, a $^{22}$Na source was introduced, simulating two 511 keV gammas back-to-back in all directions. These studies considered the ideal case of photoelectric events only and showed that a very good CRT of down to 70 ps FWHM can be achieved using VUV sensitive SiPMs with a PDE of 20$\%$ \cite{Gomez-Cadenas:2016mkq}. In addition, making use of Cherenkov light, a CRT of 30-50 ps can be obtained using very fast photosensors, sensitive to wavelengths down to 300 nm, such as microchannel plates \cite{Gomez-Cadenas:2017bfq}.

\section{Monte Carlo studies of one ring}

The following studies have been performed simulating a continuous ring without separated cells in order to reduce border effects in light distribution. The possibility of two instrumented faces was rejected because of the complexity in the mechanics required for the data acquisition and the cost of the device due to the number of channels. Some simulations were done in order to study the results in spatial and temporal reconstruction and the energy resolution for different pitches and dimensions of the sensors using only the external instrumented face. The coordinate system used is cylindrical: $R$ is the radial coordinate, $\phi$ is the angular coordinate and $Z$ is the height of the ring.

From the MC true information, the position of the deposited energy in LXe is obtained for both photons, and from the photoelectrons detected by the sensors the reconstruction of the same position is performed. The difference between both distances provides the spatial resolution, that is, the precision in identifying the position of a single-vertex interaction. In the case of the radius, if only one face is instrumented, since all sensors are located in the external ring, the depth of interaction point is extracted from a map containing the true $R$ information and the standard deviation in the $\phi$ coordinate of the sensors that detected some charge. The reconstruction of $Z$ and $\phi$ coordinates is achieved using the classical barycenter algorithm.

In the first attempt, the ring was equipped with 4512 SiPMs with active areas of 3 $\times$ 3 mm$^{2}$ covering the whole surface at a pitch of 4 mm. The radial dimensions were 15 cm of internal radius, 3 cm depth and 64 mm height. The obtained values for spatial resolution were (1.21 $\pm$ 0.07) mm FWHM and (0.79 $\pm$ 0.01) mm FWHM for $R$ and $Z$, respectively. For the angular coordinate a result of (0.22 $\pm$ 0.12) deg FWHM was achieved, which in terms of the radial coordinate is (0.57 $\pm$ 0.03) mm - (0.69 $\pm$ 0.04) mm.

Later, a ring with the same dimensions in radius but 112 mm in height and different pitch and sensor coverage, was studied. The distance between SiPMs was 7 mm, their active area was 6 $\times$ 6 mm$^{2}$ and the total number of sensors was 2576. The spatial resolution obtained was (1.03 $\pm$ 0.06) mm FWHM for radius, (0.70 $\pm$ 0.04) mm for $Z$ and (0.27 $\pm$ 0.02) deg FWHM for $\phi$, which is equivalent to (0.70 $\pm$ 0.04) mm - (0.84 $\pm$ 0.05) mm.
 
 The energy resolution $E$ of a liquid xenon scintillator depends on the variations in light collection due to the detector geometry $E_{g}$, the statistical fluctuation in the number of photoelectrons from the SiPMs $E_{s}$, the fluctuations due to electron-ion recombination  $E_{r}$ and the intrinsic resolution from liquid xenon scintillation light $E_{i}$, and can be described by: 
 
 \begin{equation}\label{energyRes}
 E^{2} = E_{g}^{2} + E_{s}^{2} + E_{r}^{2} + E_{i}^{2}.
 \end{equation}
 
 The first two components can be extracted from the MC while the contribution of the intrinsic terms ($E_{r}$ and $E_{i}$) has been measured and constitutes a factor of 14$\%$ FWHM \cite{Ni:2006zp} for 511 keV gammas. The energy resolution obtained from the MC for the case of 4 mm pitch and 3 $\times$ 3 mm$^{2}$ was (12.7 $\pm$ 0.2)$\%$ FWHM centered in the photopeak, while for the case of 7 mm pitch and 6 $\times$ 6 mm$^{2}$ area it was (9.6 $\pm$ 0.1)$\%$ FWHM centered in the photopeak.
 
 The CRT is calculated as the variance of the $\Delta t$ distribution, that is given by the following equation.
 
 \begin{equation}\label{timeRes}
 \Delta t \equiv \frac{d_{d}}{c} = \frac{1}{2} (t_{1} - t_{2} - \frac{\Delta d_{g}}{c} - \frac{\Delta d_{p}}{v_{p}}).
 \end{equation}
 
\begin{figure}[!htb]
 	\centering
 	\includegraphics[width=9.2cm]{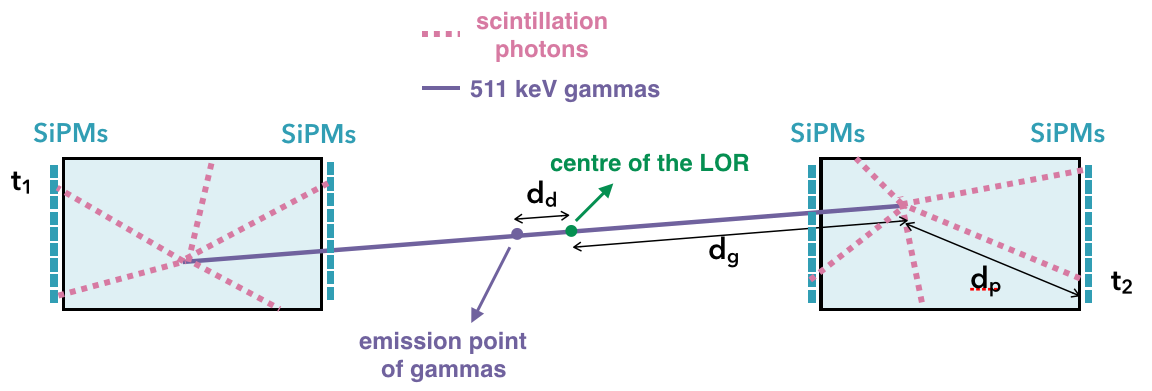}
 	\caption{Scheme of the transversal section of the ring showing the LOR of the 511 keV gammas and the scintillation photons detected by the SiPMs.}
 	\label{fig:CRT}
 \end{figure}

 As Figure \ref{fig:CRT} shows, $t_{1}$ and $t_{2}$ are the times of the first photoelectron recorded in each section, $d_{g}$ is the distance between the geometrical centre of the system and the interaction point of the 511 keV gamma, $d_{d}$ is the distance between the geometrical centre of the system and the emission point of the gammas and $d_{p}$ is the distance that the scintillation VUV photon covers between its emission and detection point.
 
 It is relevant to mention that neither the time fluctuations of the sensors nor the jitter of the front-end electronics are included in the simulations. For a discussion of the effect of jitter and electronics see Reference \cite{Gomez-Cadenas:2016mkq}. Only a PDE for the sensors of 20$\%$ and the refraction index of LXe are considered. However, in the calculation of CRT the refraction index of LXe is assumed constant, with a value of 1.7, which is the average of the values it takes in the scintillation wavelength range.
 
 The CRT obtained is (36.17 $\pm$ 1.52) ps FWHM and (34.14 $\pm$ 0.98) ps FWHM for 4 mm pitch and 7 mm pitch, respectively.
 
The importance of these results lies in the fact that an increase in the pitch of the sensors is not critical if it is accompanied by an increase in the size of their active area. Therefore, we can afford having less sensors, which means a reduction of nearly half the number of channels and implies less feedthroughs for extracting the information. It highly simplifies the mechanics and reduces the price of the device. 

However, in these results large tails appear in the resolution of the features described above, especially in $Z$, due to border effects in the detector. 

\begin{figure}
	\centering
	\begin{subfigure}[b]{0.6\textwidth}
		\includegraphics[width=0.8\textwidth]{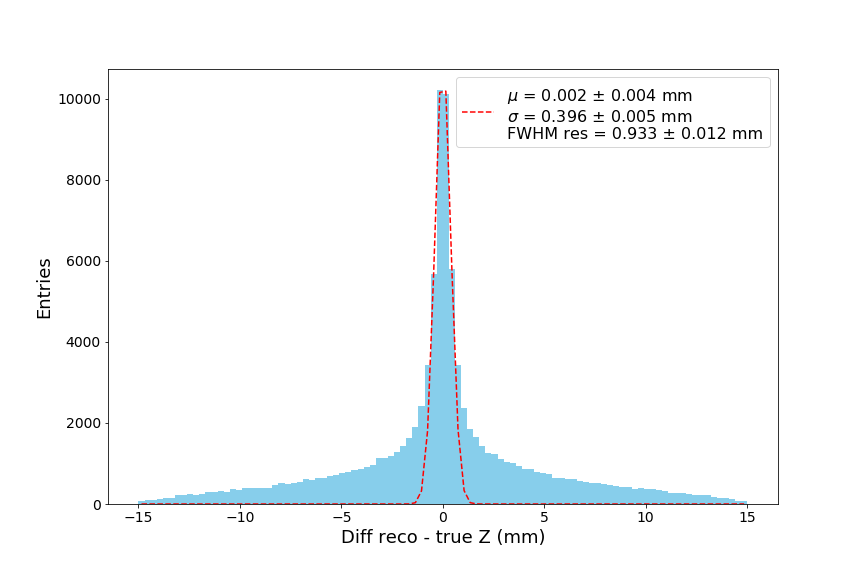}
	\end{subfigure}
	
	\begin{subfigure}[b]{0.6\textwidth}
		\includegraphics[width=0.8\textwidth]{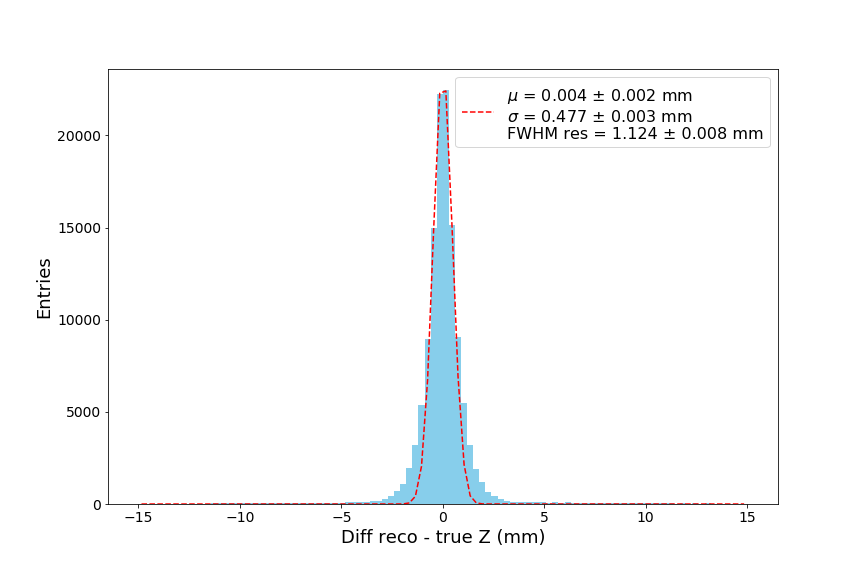}
	\end{subfigure}

	\caption{Resolution in $Z$ coordinate for the ring with $Z$ $\leq$ 8 cm (top) and $Z$ $\leq$ 80 cm (below).}
	\label{fig:plot_diff_z_th2_tiles}
\end{figure}

We explored the performance of a ring with large $Z$ in order to assess the intrinsic potential of a truly continuous detector. The pitch used in this geometry was around 5 mm and the obtained results with high $Z$, around 80 cm, were compared to a ring of height 8 cm. Plots of the resolution in the $Z$-coordinate for both ranges examined are given in Figure \ref{fig:plot_diff_z_th2_tiles}, showing the significant reduction in the number of events lying outside the gaussian if the height of the ring is increased. A behavior similar for both ranges is observed in the other coordinates, thus having the same results as the previous studies.

\section{PETALO readout}

The aim of the readout in PETALO is to send as much information as possible to increase the detector performance and efficiency. Considering the high number of channels, a commercial IC front-end (ASIC) is needed and the PETsys Time-of-Flight PET ASIC is the device chosen for that purpose \cite{PETsys}. The PETsys High Performance TOFPET2 ASIC is a new 64-channel chip for the readout and digitization of signals from fast photon detectors in applications where a high data rate and fast timing are required.

Several tests with this ASIC and the different sensor sizes and manufacturers (Hamamatsu, FBK) are being performed in order to optimise and study the response of the system at low temperatures. 

After the ASIC, an image compression technique compatible with spatial fragmentation should be introduced to reduce the output data, which could be either wavelet filter analysis or artifical neural networks \cite{readout}.

\section{PETALO first prototype: PETit}

\begin{figure}[!htb]
	\centering
	\includegraphics[height=2.3cm]{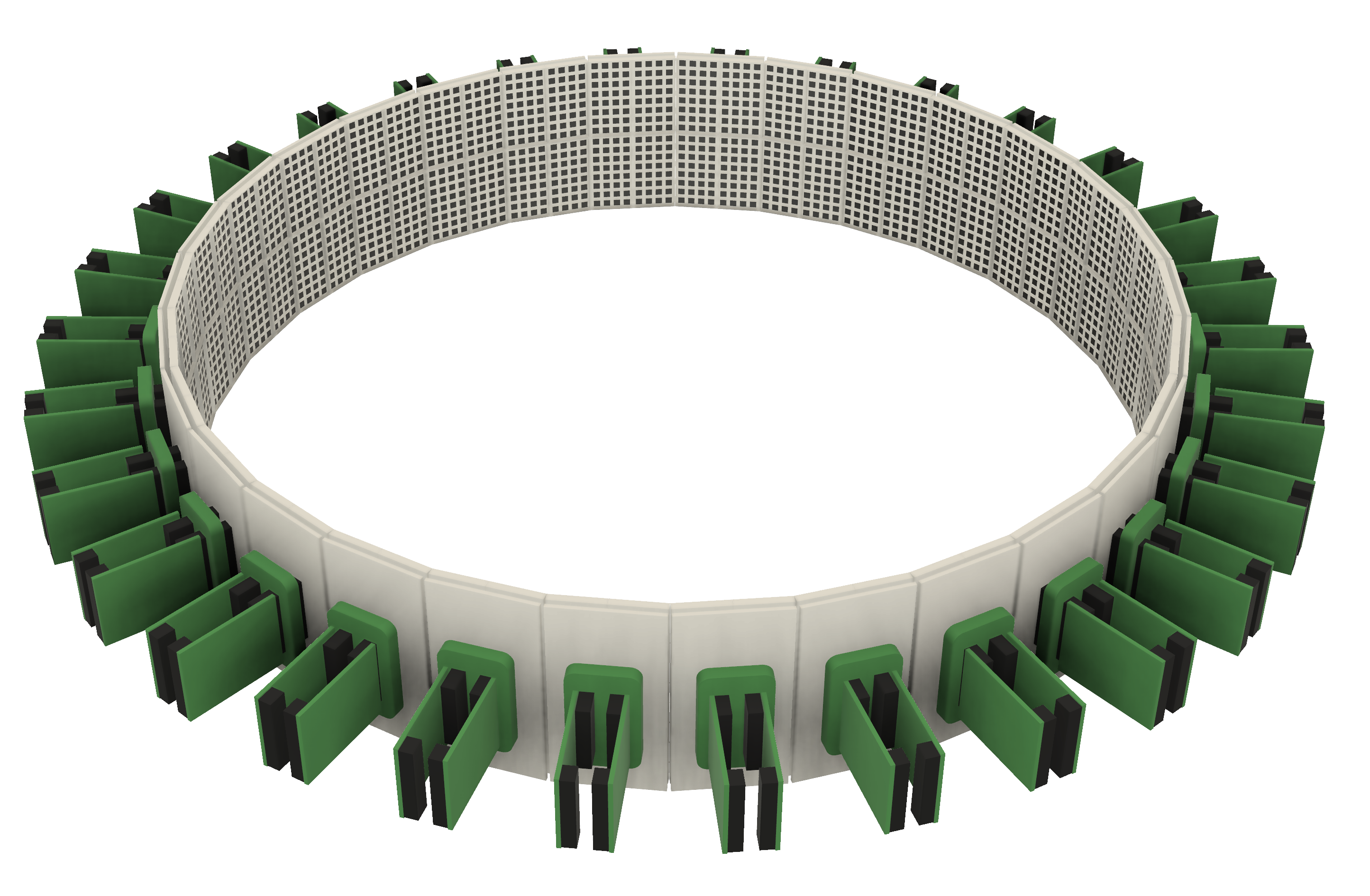}
	\includegraphics[height=2.2cm]{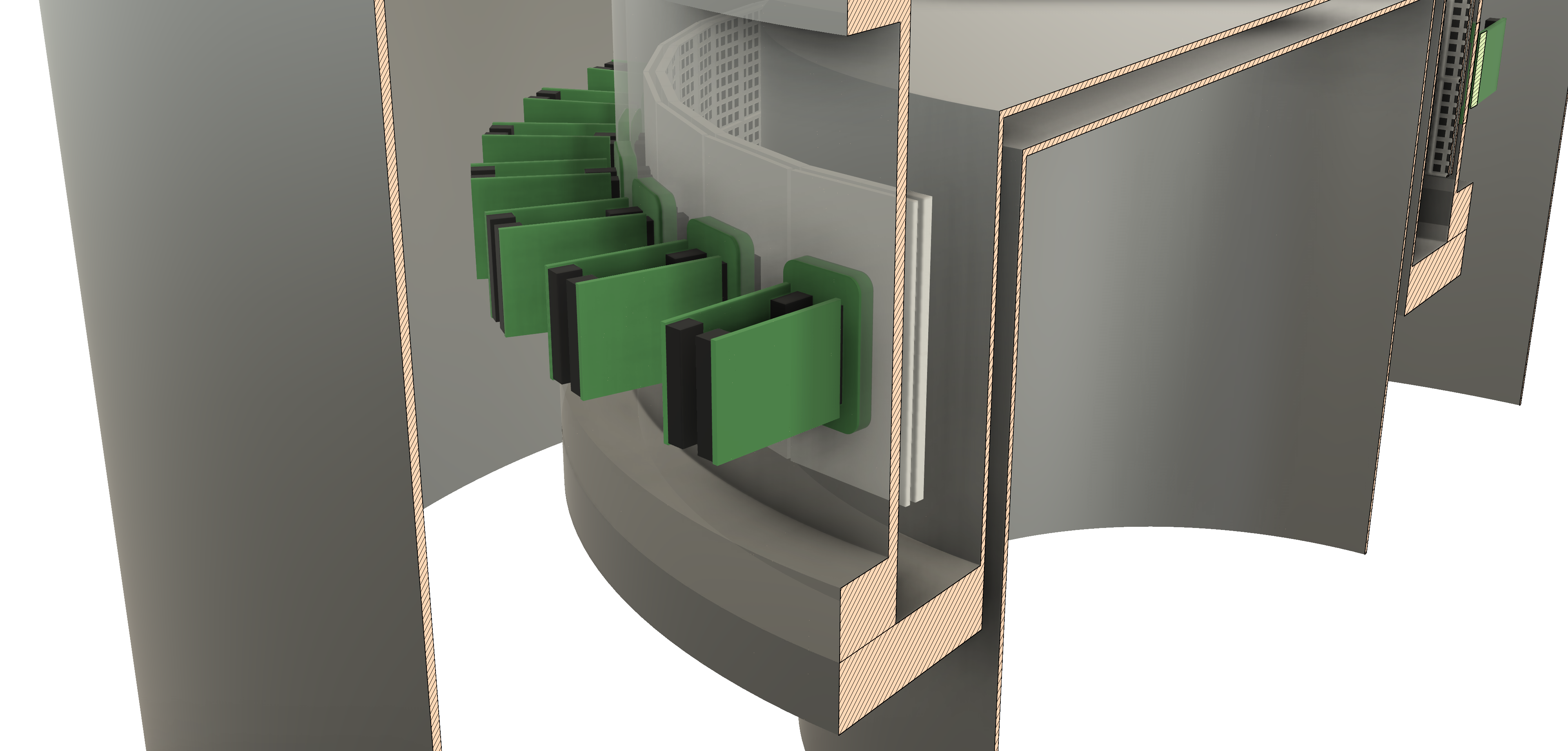}
	\caption{Simulation of the external layer of the sensors with their feedthoughs (left) and the ring placed inside the cryostat (right).}
	\label{fig:PETIT_RINGv23_v8}
\end{figure}

\begin{figure}[!htb]
	\centering
	\includegraphics[width=7.5cm]{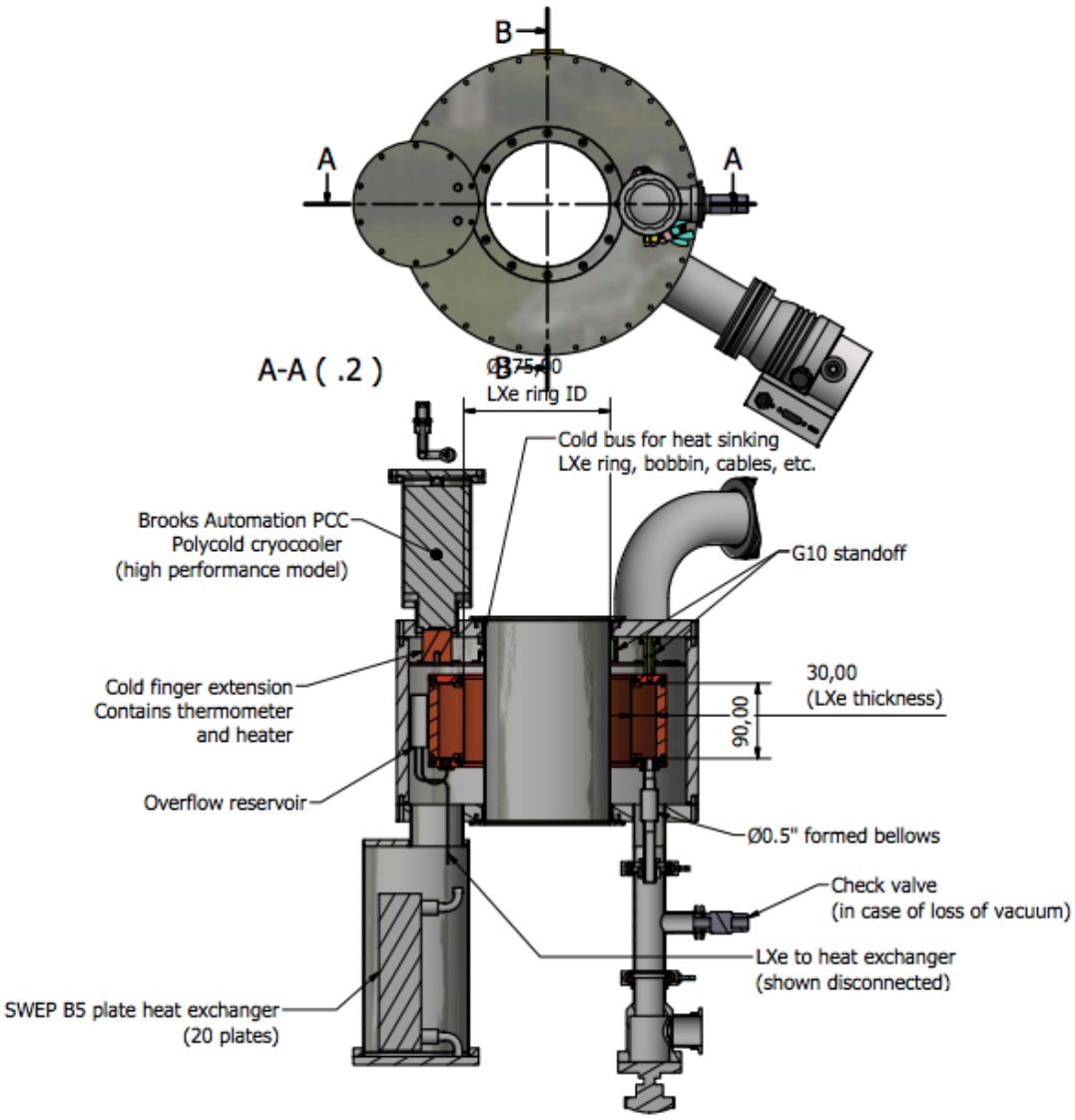}
	\caption{Preliminary design of the cryostat.}
	\label{fig:cryostat}
\end{figure}

PETit is the first PETALO prototype and it is currently in the design phase. It consists of a thermally conductive ring for liquid xenon of 30 cm diameter, 3-4 cm depth and with only the external lateral area instrumented with VUV sensitive SiPMs (Figure \ref{fig:PETIT_RINGv23_v8}). It is placed inside a high performance cryostat, the sketch of which is shown in Figure \ref{fig:cryostat} for a small ring. 

The ring will be filled with LXe, leaving the interior of the cryostat as an empty space to place the calibration sources or the corresponding phantom. PETit will start its first data taking at the beginning of 2020.

\section{Conclusions}

PETALO is a novel concept in PET scanning designed to maximize TOF performance. It combines the physical properties of scintillation and Cherenkov light of liquid xenon with the good PDE and low dark noise at low temperatures of VUV SiPMs for readout. Moreover, it is a truly continuous ring, with the benefits of a uniform response, high efficiency of light capture and minimal geometrical distortions, thus providing good energy, spatial and temporal resolution.

In this work we have studied the resolution of a ring of 15 cm radius and 3 cm depth with a single layer of sensors on the outer wall of the detector, having only one instrumented face simplifies the electronics and affects the spatial and energy resolution very little.

Different pitches and coverage for the SiPMs have been simulated and it has been demonstrated that larger pitches can be used if the active area of sensors is larger as well, therefore allowing for the simplification of the mechanics and costs.

Finally, it has been shown that increasing the height of the ring significantly improves the resolution in the $Z$-coordinate.

PETit is the first demonstrator of the PETALO concept. This prototype is a continuous ring of 15 cm internal radius whose goals are to obtain good measurements for energy, spatial and temporal resolution and to demonstrate that a complete image reconstruction of a phantom is possible. Suqsequently, the technology will be scaled to larger sizes up to human body PET.

\section{Acknowledgments}
 
  The author would like to thank the support of the following institutions and agencies: the European Research Council (ERC) under Starting Grant 757829-PETALO; the Spanish Ministry of Economy and Competitiveness under project FPA2016-78595-C3-1-R and the Generalitat Valenciana of Spain under grant PROMETEO/2016/120.

\section*{References}

\bibliography{mybibfile}

\end{document}